\documentstyle[12pt,aasms4]{article}
\newbox\grsign \setbox\grsign=\hbox{$>$} \newdimen\grdimen \grdimen=\ht\grsign
\newbox\simlessbox \newbox\simgreatbox
\setbox\simgreatbox=\hbox{\raise.5ex\hbox{$>$}\llap
     {\lower.5ex\hbox{$\sim$}}}\ht1=\grdimen\dp1=0pt
\setbox\simlessbox=\hbox{\raise.5ex\hbox{$<$}\llap
     {\lower.5ex\hbox{$\sim$}}}\ht2=\grdimen\dp2=0pt
\def\simgreat{\mathrel{\copy\simgreatbox}}

\begin{document}

\title{The Polarizing Power of the Interstellar Medium in Taurus}
\author{H\'ector G. Arce\altaffilmark{1,}\altaffilmark{2}, Alyssa A.
Goodman\altaffilmark{1,}\altaffilmark{3}, Pierre
Bastien\altaffilmark{4}, Nadine
Manset\altaffilmark{4} and Matthew Sumner\altaffilmark{1,}\altaffilmark{5}}

\altaffiltext{1}{Harvard-Smithsonian Center for Astrophysics, 60 Garden Street, Cambridge, MA 02138.}
\altaffiltext{2}{National Science Foundation Minority Graduate Fellow}
\altaffiltext{3}{National Science Foundation Young Investigator}
\altaffiltext{4}{Universit\'e de Montr\'eal, D\'epartement de Physique,
and Observatoire du Mont M\'egantic, C. P. 6128, Succ. Centre-ville,
Montr\'eal, P.Q., H3C 3J7 Canada.}
\altaffiltext{5}{Current Address: California
Institute of Technology, Department of Physics, Pasadena, CA 91125.}

\begin{abstract}
We present a study of the polarizing power of the dust in cold dense
regions (dark
clouds) compared to that of dust in the general interstellar medium (ISM).
Our study uses new
polarimetric, optical, and spectral classification data for 36 stars to
carefully study the
relation between polarization percentage ($p$) and extinction ($A_V$) in
the Taurus dark
cloud complex.   We find two trends in our $p-A_V$ study: (1) stars
background to the warm ISM show an increase in $p$ with
$A_V$; and
(2) the
percentage of polarization of
stars background to cold dark clouds does not increase with extinction.
{\it We detect a
break in the $p-A_V$ relation at an extinction $1.3 \pm 0.2$ mag,
which we expect corresponds to a set of conditions where the polarizing
power of the dust
associated with the Taurus dark clouds drops precipitously. This breakpoint
places important
restrictions on the use of polarimetry in studying interstellar  magnetic
fields.}
\end{abstract}

\keywords{dust, extinction --- ISM: clouds --- ISM: magnetic fields ---
polarization} 

\section{Introduction}

The polarization of background starlight has been used for nearly half a
century
to probe the magnetic field  direction in the interstellar medium (ISM).  The
observed polarization  is believed to be caused by dichroic  extinction of
background starlight passing through concentrations of aligned elongated dust
grains along the line-of-sight. Although there is no general consensus on
which is the dominant grain alignment mechanism (Lazarian, Goodman \& Myers
1997), it is  generally believed that the shortest axis of the ``typical''
elongated grain tends to be become  aligned to the local magnetic field. For
this orientation, the observed polarization vector is parallel to
the plane-of-the-sky projection of a line-of-sight-averaged magnetic field
(Davis \& Greenstein 1951).

The line-of-sight averaging inherent in background starlight polarimetric
observations can make interpretation of the polarization produced by different
field orientations and/or several independent dust clouds very complicated.
Nonetheless, it was thought that if lines of sight with just one localized very
dusty region (such as a dark cloud) between us and a background
star could be found, surely the polarization would reveal the field associated
with that dusty region. However, recent studies in the Taurus region
(Goodman et al. 1992; Gerakines et al. 1995) and other parts of the 
sky (Creese et al. 1995; Goodman et al. 1995)
have uncovered
substantial evidence to show that dust inside cold dark clouds has lower
polarizing power
than dust in the general warm ISM.
This means that the polarization of the
light from a background
star is a non-uniformly {\it weighted} line-of-sight average of the projected
plane-of-the-sky field, and that grains in cold dark clouds are
systematically down-weighted.
The ultimate implication of this down-weighting is that above some
(column?) density
threshold, the polarization of background starlight gives no information
about the magnetic
field in dark clouds.  It is the goal of this Letter to find and physically
describe this
threshold.

The evidence that grains in cold dark clouds are inefficient polarizers of
background starlight
is multi-faceted.  Eight years ago, Goodman et al. (1990) found that the
smooth large-scale
patterns apparent in polarization maps of dark cloud complexes (e.g. Vrba
et al.
1976; Vrba et al. 1981; Moneti  et al. 1984; Whittet et al. 1994) do not
systematically change
in response to the large density enhancements represented by the dark
clouds.  After this
realization, it was hypothesized that perhaps optical polarimetry was
incapable of seeing
field changes which might occur only in the high-density, optically opaque,
interiors of dark
clouds.  So, near-infrared polarimetry, which can probe the optically
opaque cloud interiors
was undertaken.  The near-infrared observations showed that the mean
direction and dispersion
of the polarization vectors are virtually {\it identical} in the cloud
interiors and their
peripheries (Goodman et al. 1992; 1995).  Thus, the presence of cold dark
clouds still
appeared to have no geometric effect on the polarization maps, implying
either that: 1) the
field is truly unaffected by the cloud; or 2) that background starlight
polarimetry is somehow
insensitive to the field in dark clouds. Polarization-extinction relations
provide the best
discriminant between these hypotheses.  For grains of constant polarization
efficiency,
$p$ should rise with
$A_V$.  Using the near-infrared observations, Goodman et al. (1992, 1995)
find that the {\it
percentage of polarization does not rise with extinction} in cold dark
clouds.  The simplest
interpretation\footnote{Note that increased field tangling inside dark
clouds cannot explain
the near-infrared polarimetric observations for two reasons. 1.) The
dispersion in the
distribution of position angle does not increase in the cloud interior
(near-IR observations)
relative to the periphery (optical observations).  And, 2.) while it is true
that the
slope of a
$p-A_V$ relation will diminish due to field tangling, it will remain
positive even for highly
tangled fields if all grains polarize equally well (see Jones 1989; Jones
et al. 1992).}  of
this result is that dust in dark clouds adds plenty to the observed
extinction, but has very
little ``polarizing power" and so adds only a very small fraction to
the observed net polarization.
A number of factors, including poor grain alignment, grain growth, and/or
changes in grain
shape or composition, could be responsible for the low polarization
efficiency exhibited by
dust grains in cold dark clouds (see Goodman et al. 1995).
Regardless of which factor(s) cause(s) the low polarization efficiency
exhibited by by dust in dark clouds, the fact is that 
background starlight polarimetry does not reliably reveal the
magnetic field {\it
in} dark clouds.

Based on the
near-infrared studies, we expect that the fraction of grains with high
polarization
efficiency  is relatively constant in the lower-density warm ISM, but drops
precipitously in
dark clouds.  Therefore, we hypothesize that a breakpoint in the
$p-A_V$ relationship might exist at the dark clouds' ``edges", which
previous studies in the
near-IR   (Goodman et al. 1992; 1995; Gerakines et al. 1995)
could not detect, due to their
inability to measure low
$A_V$'s accurately enough. In this Letter, we present our attempt to
carefully study the
$p-A_{V}$ relation near dark clouds, and thus offer a set of guidelines
as to where the polarization maps can be taken as
faithful
representations of the magnetic field projected onto the plane of the sky,
and where they
cannot.

\section{Data}

Our observing strategy consists of three parts. First we obtain CCD images
(using
$B$, and $V$ broad-band filters) of two 10 arcmin by $\sim$ 5 deg ``cuts''
(see Figure 1)
through the Taurus dark cloud complex. Second, we measure the spectrum of
94 stars along the
cuts in order to determine their spectral types. Using the multi-color
photometry and the
spectral types, we derive an extinction and distance to each of the stars,
using the
relation $A_V=R_V E_{B-V}$, with $R_V=3.1$  (Savage \& Mathis 1979; Vrba \&
Rydgren 1985).
Third, we measure the polarization of as many of the stars for which we had
extinction
values based on the photometry and spectral classification as time permits.
(The
polarimetry is by far the most time-consuming of the three steps.)

The cuts shown in Figure 1 pass at approximately right angles through two
well known
highly-elongated dark clouds in the Taurus complex, L1506 and B216-217,
both of which are
thought to lie 140 $\pm$ 10 pc from the Sun (Kenyon et al. 1994). 
We chose this
orientation for the cuts in
order to insure both: 1) a sample of stars with a large dynamic range in
extinction; and 2)
that many of the high $A_V$ values measured would be produced by a single
localized dark cloud
along the line of sight.  We attempted to exclude foreground stars by not
selecting stars
which appear relatively bright, and the measured spectroscopic parallax
distances (see Arce \&
Goodman 1998, hereafter AG98) confirm that we largely succeeded in doing
so. (Our stellar sample only has three stars with distances less than 150 pc.)  In total, the polarization of 31 stars was measured.  In addition, we used previously published data to
obtain color
excesses for two stars (Kenyon et al. 1994) and  polarization information
for 5 stars (Goodman
et al. 1990).

The broad band images were obtained using the CCD camera on the 
Smithsonian Astrophysical Observatory (SAO) Fred Lawrence
Whipple Observatory (FLWO) 1.2-meter telescope on Mt.~Hopkins, Arizona.
The stellar spectra were obtained using the SAO FAST spectrograph on the FLWO 1.5-meter telescope. The observations were carried out
during the Fall trimesters of 1995 and 1996. For information on the 
reduction of the photometric and spectroscopic data see AG98.   
Employing an analysis method similar to that of Wood et al. (1994), we also
created an
extinction map of Taurus, with 5\arcmin \/ resolution, using co-added images of
flux density from the
Infrared Astronomical Satellite (IRAS) Sky Survey Atlas (see AG98).

The polarization observations were carried out at the Mont M\'egantic
Observatory, using the
Beauty and The Beast two-channel photoelectric polarimeter (described in
Manset \& Bastien
1995) on the 1.6-meter Ritchey-Chr\'etien telescope, in February 1996,
December 1996 and
February 1997. We used a Schott RG-645 filter which, combined with the 
photomultiplier response, gives a broad passband
of 2410 \AA,
centered at 7660 \AA.  A small instrumental polarization of  $\sim 0.1\%$
was measured and
subtracted from the observations. The polarization scale was determined by
observing a bright
star and using a calibration prism which gives essentially
100\% polarized light.
The origin of the position angles was determined by observing standard
polarized stars. The
uncertainty in the polarization angle ($\theta$) is given by the formula
$\sigma(\theta) =
28\arcdeg.65[\sigma(p)/p]$ \/ (Serkowski 1974).

Table 1 summarizes the relevant polarization and extinction values derived
from all of
the data collected and assembled.

\section{Results}
Plots of polarization vs. extinction are shown in Figure 2. There seem to
be two different
trends in the $p-A_V$ plots.  Most stars with $A_{V} \lesssim 1.3$ mag
follow a trend in which
their percentage of polarization increases with extinction. But many stars
with $A_{V} \gtrsim
1.3$ mag keep a more or less constant $p$ with extinction. Upon overlaying the
stars' positions
on an extinction map (Figure 1), it becomes apparent that several of the
stars which have high
extinction, but low polarization are very near peaks in the optical depth
(i.e., dark clouds).
On the other hand, the positions of most of the other stars which follow an
increasing $p$
with increasing $A_{V}$ tend to lie either on places were the optical depth
is small (like the
southernmost stars) or on small, randomly occurring, high extinction
patches, far away from the
optical depth peaks.

From the results of previous studies we can easily guess that the $p-A_V$
relation for stars
that are background to cold dark clouds might be different from that of
stars background to the
general (lower-density) ISM.  Thus, it makes sense to attempt to 
systematically separate
our stellar sample
into stars that are and are not ``background to dark clouds," in order to
study the potentially
different
$p-A_V$ relations. Visually inspecting the positions of the stars
superimposed on an
extinction map (see Figure 1) is not an adequately quantitative method of
reliably achieving
this separation.  
Our definition of a star ``background to a dark cloud" is
one which lies
close to an extinction peak {\it and} is primarily extinguished by dust in
dense regions at a
single distance.  With this definition in mind, we use two different means of
differentiating  between the two ``types'' of stars.

The first method uses the IRAS $A_V$ map. The three main extinction ridges
which cross the
slice of Taurus shown in Figure 1 are  L1506, B216-217 and the region Wood
et al.
(1994) name IRAS core Tau M1. 
For each star which lies near one of these ridges, we
plot $A_V$ along a
line on the plane of the sky defined by the normal to the nearest ridge
from the star.
For each of these $A_V$ vs. offset profiles, we fit a gaussian to the
resulting extinction profile and then
measure the distance from the $A_V$ peak to the star's position, in units
of half width at
half maximum (HWHM).  The results of this procedure are shown in Table 1.
We then divide the sample into two groups: stars that  appear close to the
optical depth peaks
in projection and stars that appear  far. We experimented with different
distances as the
boundary between the two groups, and  decided to finally adopt 1.6 HWHM as
such. With 1.6 HWHM
as the dividing  value we obtain two clear
$p-A_V$ trends in the sample, while  having the dividing distance between
close and far stars
from the dark cloud extinction  peaks as far as possible.

The second method of separating the stars uses the fact that the observed
extinction can be
thought of as a sum of the extinction due to the dense regions of the dark
cloud and the
extinction due to the dust foreground and background to the dark cloud.
Using the same extinction profiles determined from the IRAS map in the
first method, we
assume that by subtracting a baseline from the gaussian we can eliminate
the extinction due
to the dust foreground and background to the cloud, and that the
baseline-subtracted
gaussian gives an estimate of the extinction due to the dense region of the
cloud.  The
value of the baseline-subtracted gaussian at the stars' projected position
can then be
compared to the total value of IRAS $A_V$ at the same position, giving an
estimate of the
percentage of extinction due to the dark cloud itself. We say that a star
is ``background to a
dark cloud" if more than 20\% of the IRAS $A_V$ at the star's projected
position
is due to the baseline-subtracted gaussian. A potential problem with this
method is 
that the dark cloud does not necessarily has to have a gaussian
profile---it could be
gaussian-like, with broad wings. If this is so, then by subtracting a
baseline, one is
potentially subtracting a substantial part of the extinction due to the
dark cloud.

We classify a star
as being
``background to a dark cloud" if it satisfies the background star criteria
for both of the
two different methods (see Table 1).

{\it Figure 2 clearly shows that $p$ increases with $A_V$ for stars that
were classified
background to the ISM, while $p$ is roughly constant with $A_V$ for stars
classified as being
background to dark clouds.}
After making
linear least-square fits, weighted by the uncertainty in $p$, to the two
different groups, we
find that the trends cross at $A_V=1.3$ mag. The errors in the fitted line
coefficients imply
an uncertainty of
$\pm$ 0.2 mag for this intersection point. Thus, we conclude that there is
a break in the
$p-A_V$ relation for this region of Taurus at $A_V = 1.3 \pm 0.2$ mag.

We believe that choosing a constant value  of $R_V$ to derive the
extinction to each star does
not adversely affect our results. Studies have shown that denser regions
usually harbor bigger
grains (Whittet \& Blades 1980) and bigger grains imply larger values of
$R_V$ (Cardelli et
al. 1989), which in turn imply a greater value of $A_V$ for a given value of
$E_{B-V}$. Thus we
expect that if there are changes in $R_V$ towards different lines of sight,
then $R_V$ should
be larger for lines of sights that pass through dark clouds. If this is so
then the stars
observed through these lines of sight would have more extinction than
originally calculated.
These points would then shift to the right in our $p$ vs. $A_V$ plot  in
Figure 2. If this
happened, the break in the $p-A_V$ relation would just be more pronounced
and clear.  We are
relatively certain that the value of $R_V$ does not vary in a systematic
way for stars
background to the general ISM, since their color-excess-determined
extinction agrees very well
with the other methods of obtaining extinction (see AG98).

\section{Analysis and Discussion}

The two trends we find in our data---rising $p$ with $A_V$ and roughly
constant $p$ with
$A_V$---have each been observed in previous studies.   But these trends
have never been
observed together, in the same region, as clearly distinct trends before.
In fact, most
studies find very large scatter in their $p-A_V$ plots, and a
clear linear
correlation in the
$p-A_V$ relation for stars background to warm dust---our first trend---is
only observed in
a few studies, such as Wilking et al. (1979) or Guetter \& Vrba (1989).
Theoretically, one should
observe this linear correlation if along all lines of sight each grain
polarizes light
equally well, and there are no big changes in the field orientation. Most
probably this is
the case in these two studies since most of their lines of sight pass
through similar
environments, in the small region of the sky that they studied. The second
trend in our
data---a slow or no rise in $p$ with $A_V$---has been observed in near-IR
polarimetric studies
of stars background to cold dark clouds (Goodman et al. 1992; 1995).

In general, most studies of the $p-A_V$ relation find that there is a fuzzy
upper
bound to the amount of polarization possible for a particular extinction
and that observed
points lie anywhere below this upper limit in the $p-A_V$ plane (Hiltner
1956; Vrba et al.
1976; Vrba et al. 1981; Moneti et al. (1984); Vrba et
al. 1993; Whittet et
al. 1994; Gerakines et al. 1995).
  Such results can easily be explained by
allowing the
polarization efficiency of the grains and/or the field orientation with
respect to the plane
of the sky to vary from one line of sight to another (Hiltner
1956). Polarization studies of regions
 near dark clouds most probably sample lines of
sight passing through a variety of different physical environments
---lines of sight with different 
dust and gas densities, which 
likely contain grains with different average polarizing efficiencies.
In most cases, the samples in these studies are not separated depending on 
the characteristics
of the line of sight, thus the upper-envelope behavior exhibited by
their $p-A_V$ relations is expected. 
Moreover, many of these studies also 
find a decrease in polarization
efficiency ($p/A$) with increasing extinction which is
likely due to the fact that the highly extinguished stars often lie behind dark
clouds---where dust is likely to have a lower polarization efficiency.

A break in the $p-A_{V}$  relation can be predicted for stars background to
quiescent dark
clouds by many different theories. Most recently, Draine \& Weingartner
(1996) studied the
effect of radiative torques due to the anisotropic illumination of
helical grains. They came
to the conclusion that, in the average ISM, diffuse clouds, and warm star
forming  regions,  such  a process is able to produce rotational velocities
higher than 
the ones produced by suprathermal mechanisms studied previously (e.g.
spin-up caused by H$_2$
formation on grains; Purcell 1979). But, on the other hand, they conclude
that inside
quiescent dark clouds the radiative torques are  unimportant due to the
weakened radiation
fields, and  higher matter density. They explicitly state that they expect
grain alignment
near the cloud surface, but not at depths of
$A_{V} \gtrsim 2$ mag. Moreover, a quantitative study of six types of
alignment  mechanisms in
the dark cloud L1755 was done by Lazarian,  Goodman \& Myers (1997). Using
data from the
literature  they studied the joint action of different  alignment
mechanisms in the outer,
intermediate and inner  regions of the cloud. They came to the conclusion
that all the  major
mechanisms fail to produce alignment in the inner  and intermediate regions
(which they assume
to  have $A_{V} \gg 1$), while grain alignment is efficient in the  outer
regions ($A_{V}
\lesssim 1$). The value we obtain for  the break in the $p-A_{V}$ relation,
$A_V = 1.3 \pm
0.2$ mag, is consistent with the these two studies.
These results are not necessarily inconsistent with the observed polarization
in the 3\micron \/ ice feature (Hough et al. 1988).
Polarization in the 
ice feature is only in
stars background to star-forming clouds (HCl 2,
OMC-1) which are more likely warmer than the filamentary cold dark clouds
in our study, and hence can harbor grains with higher polarization efficiency
(see Figure 7 in Goodman et al. 1995).

\section{Conclusion}

The breakpoint in the $p-A_{V}$ relation places important restrictions on
the use of
polarimetry in studying interstellar magnetic fields. Since the
polarization efficiency of the
dust inside dark clouds is very low, most of the polarization 
observed for lines
of sight which pass through these extinction peaks is
not due to the dark cloud;
it is due to dust background
and foreground to
the cloud. Hence, one should not use background starlight polarimetry to
map magnetic fields
inside dark clouds. With the results of this study we can quantify the word
``inside." In
regions like Taurus, {\it it is safe to interpret the polarization of
background starlight
as a representation of the plane-of-the-sky projected magnetic field up to
the 1.3
$\pm$ 0.2 mag ``edge'' of the dark cloud}.  In other words the linear
relation between $p$ and
$A_V$ that exists in the low-density ISM breaks down for stars background
to the $\simgreat
1.3$ mag of extinction produced by a dense localized dusty region (i.e.,
dark cloud).  After
this edge polarization no longer rises with extinction, and thus cannot
reveal the field
structure in the dense region. We restate that this proscription  only
applies for stars
background to cold dark clouds, as stars background to the warm ISM have
not been shown to
exhibit such behavior.

\acknowledgements
H. G. A. gratefully acknowledges the support from the National Science
Foundation Minority Graduate Fellowship.
We would like to extend our gratitude to Scott J. Kenyon, 
Lucas M. Macri and Perry Berlind for their great help in  
the photometry analysis, spectral classification of stars, 
and acquisition of data.

\clearpage

%\section{figure captions}

%one
\figcaption[figure1.eps]{Extinction ($A_V$) map of the region under study, obtained employing an analysis method similar to that of Wood et al. (1994), 
using co-added images of flux density from the
Infrared Astronomical Satellite (IRAS) Sky Survey Atlas (see AG98).
The position of the 36 stars in our sample
is shown: $\bullet$ are stars background to the low-density
ISM; $\triangle$ are stars with a distance of less than 150 pc; 
$\Box$ are stars classified as ``background to dark clouds''. 
The lines mark the position of the two cuts.
\label{fig1}}

%two
\figcaption[figure2.ps]{Observed relation between polarization and extinction,
where $A_V=3.1E_{B-V}$. The lines are least square linear fits, weighted
by the uncertainty in $p$. The dashed line is the fit to points representing
the stars background to the low-density ISM, which gives 
$p=(0.09\pm0.06)+(3.58\pm0.13)E_{B-V}$, with a correlation coefficient of
0.68. The solid line is the fit to the points representing the stars
background to dark clouds, which gives $p=(1.61\pm0.13)+(0.03\pm0.15)E_{B-V}$,
with a correlation coefficient of 0.79.
\label{fig2}}

\clearpage

\begin{deluxetable}{ccccccc}
\scriptsize
\tablecolumns{7}
\tablecaption{Polarization Data
\label{poltab}}
\tablehead{
\colhead{Star} &
\colhead{$p$}  & \colhead{$\theta$} 
& \colhead{$A_{V}$\tablenotemark{b}} & 
\colhead{HW} & \colhead{\%\tablenotemark{d}} &
\colhead{Note\tablenotemark{e}}
\\
\colhead{Name\tablenotemark{a}} & \colhead{(\%)} & \colhead{(E of N)} &
\colhead{[mag]} &
\colhead{HM\tablenotemark{c}} & \colhead{} & \colhead{}
}
\startdata
011005 & $1.11\pm0.24$ & $79.2\pm6.2\arcdeg$ & $0.73\pm0.16$ & --- & --- & no\nl
021013 & $0.71\pm0.34$ & $56.5\pm13.7$ & $0.72\pm0.23$ & --- & --- & no\nl
021012 & $0.35\pm0.31$ & $41.0\pm25.7$ & $0.64\pm0.16$ & --- & --- & no\nl
021011 & $0.89\pm0.39$ & $35.9\pm12.5$ & $0.73\pm0.16$ & --- & --- & no\nl
031023 & $0.13\pm0.23$ & $98.3\pm51.1$ & $0.70\pm0.16$ & --- & --- & no\nl
041033 & $2.16\pm0.45$ & $41.6\pm5.9$ & $1.59\pm0.27$ & 2.3 & 14 & no\nl
041032 & $1.51\pm0.24$ & $61.4\pm4.6$ & $1.50\pm0.23$ & 1.4 & 30 & yes\nl
051041 & $1.09\pm0.14$ & $63.2\pm3.7$ & $1.08\pm0.17$ & --- & --- & no\nl
071064 & $1.33\pm0.08$ & $70.3\pm1.8$ & $1.37\pm0.15$ & --- & --- & no\nl
071062\tablenotemark{f} & $2.08\pm0.08$ & $70.4\pm1.1$ & $1.74\pm0.30$ & --- &
--- & no\nl
081075 & $1.48\pm0.25$ & $75.5\pm4.9$ & $1.65\pm0.26$ & 3.0 & 0.7 & no\nl
AG-136\tablenotemark{g} & $1.60\pm0.10$ & $83.0\pm2.0$ & $1.33\pm0.15$ & 1.5 &
27 & yes\nl
AG-133\tablenotemark{g} & $1.83\pm0.11$ & $76.0\pm2.0$ & $1.27\pm0.18$ & 2.2 & 9 & no\nl
AG-132\tablenotemark{g} & $1.37\pm0.18$ &  $7.0\pm4.0$ & $1.67\pm0.19$ & 1.6 &
20 & yes\nl
AG-102\tablenotemark{g} & $2.10\pm0.06$ & $99.0\pm1.0$ & $3.10\pm0.34$ & 0.9 &
34 & yes\nl
091084 & $0.99\pm0.65$ & $101.2\pm18.5$ & $2.27\pm0.24$ & 0.9 & 40 & yes\nl
092087 & $1.14\pm0.13$ & $96.4\pm3.2$ & $2.45\pm0.17$ & 1.6 & 21 & yes\nl
AG-105\tablenotemark{g} & $0.93\pm0.11$ & $85.0\pm3.0$ & $1.58\pm0.24$ & 2.1 & 4 & no\nl
091081 & $2.18\pm0.40$ & $28.8\pm5.3$ & $1.11\pm0.15$ & --- & --- & no\nl
101095 & $2.07\pm0.45$ & $45.6\pm6.3$ & $1.34\pm0.16$ & --- & --- & no\nl
101094 & $1.21\pm0.18$ & $55.5\pm4.3$ & $1.35\pm0.15$ & --- & --- & no\nl
S76573\tablenotemark{f,h} & $0.03\pm0.08$ & $139.5\pm57.7$ & $0.00\pm0.15$ & ---
& --- & no\nl
101091 & $2.16\pm0.22$ & $60.6\pm2.9$ & $1.57\pm0.16$ & --- & --- & no\nl
111104 & $0.46\pm0.12$ & $96.5\pm7.6$ & $1.41\pm0.15$ & --- & --- & no\nl
111101 & $1.51\pm0.08$ & $43.6\pm1.5$ & $1.80\pm0.35$ & --- & --- & no\nl
121115 & $1.39\pm0.18$ & $55.4\pm3.7$ & $1.78\pm0.22$ & --- & --- & no\nl
121113 & $2.75\pm0.32$ & $49.4\pm3.3$ & $1.45\pm0.22$ & --- & --- & yes\nl
131123 & $2.41\pm0.29$ & $60.0\pm3.4$ & $1.85\pm0.26$ & 1.2 & 42 & yes\nl
131121 & $1.44\pm0.28$ & $21.1\pm5.7$ & $1.83\pm0.23$ & 0.3 & 41 & yes\nl
141135 & $1.54\pm0.21$ & $37.9\pm3.9$ & $2.29\pm0.16$ & 0.2 & 40 & yes\nl
141134 & $1.47\pm0.09$ & $43.3\pm1.8$ & $3.73\pm0.16$ & 0.2 & 40 & yes\nl
141133 & $2.70\pm0.25$ & $28.2\pm2.7$ & $1.76\pm0.15$ & 0.9 & 23 & yes\nl
00B2.2 & $1.57\pm0.14$ & $11.1\pm2.5$ & $2.03\pm0.16$ & 1.5 & 33 & yes\nl
00B2.1 & $2.08\pm0.13$ & $17.1\pm1.9$ & $1.35\pm0.16$ & 2.6 & 1 & no\nl
TDC321 & $2.87\pm0.17$ & $12.3\pm1.7$ & $2.87\pm0.17$ & --- & --- & no\nl
S76574\tablenotemark{f,h} & $2.60\pm0.08$ & $10.2\pm1.0$ & $1.46\pm0.15$ & --- &
--- & no\enddata
\tablenotetext{a}{The stars are ordered in increasing declination, the 
coordinates are given in AG98.}
\tablenotetext{b}{The error in $A_V$ does not include the error introduced
by assuming a constant value of $R_V$.}
\tablenotetext{c}{Distance from nearest dark cloud $A_V$ peak to the star's 
position in units of half width at half maximum of the fitted gaussian.}
\tablenotetext{d}{Percentage of the star's total IRAS $A_V$ which
is due to the baseline-subtracted gaussian.} 
\tablenotetext{e}{Was the star classified as being background to a dark
 cloud?}
\tablenotetext{f}{Stars with a distance less than 150 pc.}
\tablenotetext{g}{Polarization data obtained from Goodman et al. (1990).}
\tablenotetext{h}{These are SAO stars. Their photometry, spectral
classification
and distance were obtained from Kenyon et al (1994).}
\end{deluxetable}

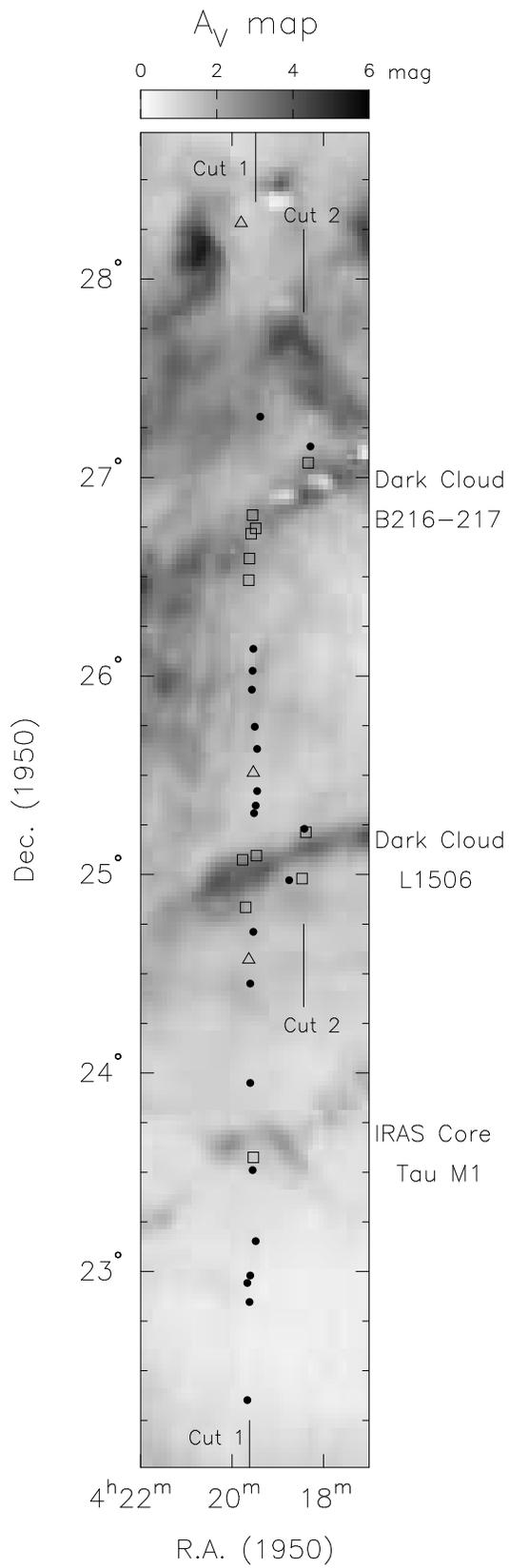
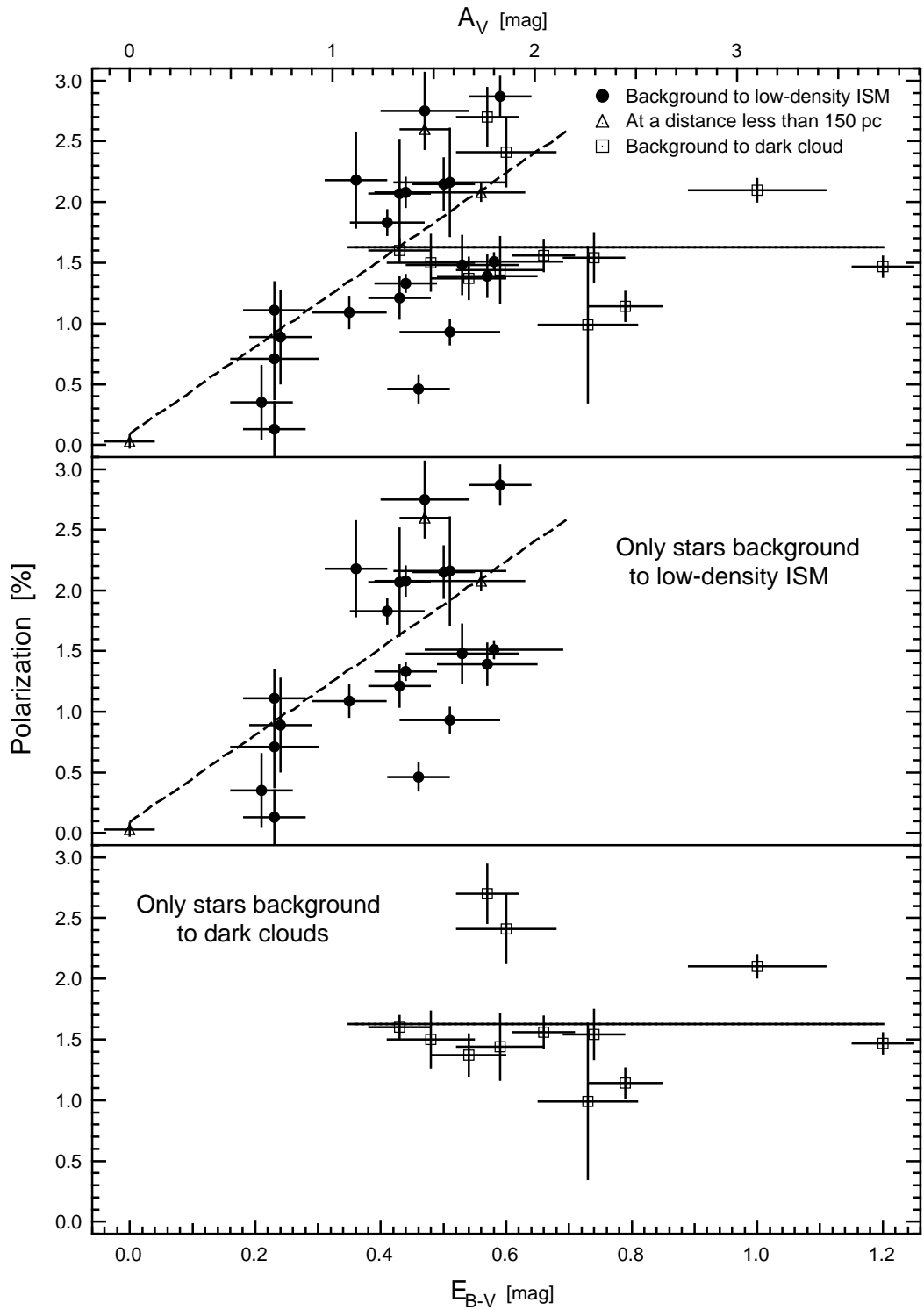
\end{document}